\documentclass[amsmath,amssymb,preprint,showpacs]{revtex4}

\usepackage{graphicx}
\usepackage[usenames]{color}

\begin{document}

\title{Optical properties of the Ce and La di-telluride charge density wave compounds}
\author{M. Lavagnini, A. Sacchetti and L. Degiorgi} \affiliation{Laboratorium f\"ur
Festk\"orperphysik, ETH - Z\"urich, CH-8093 Z\"urich,
Switzerland.}
\author{K.Y. Shin and I.R. Fisher}
\affiliation{Geballe Laboratory for Advanced Materials and
Department of Applied Physics, Stanford University, Stanford,
California 94305-4045, USA.}

\date{\today}

\begin{abstract}
The La and Ce di-tellurides LaTe$_2$ and CeTe$_2$ are deep in the charge-density-wave (CDW) ground state even at 300 K. We have collected their electrodynamic response over a broad spectral range from the far infrared up to the ultraviolet. We establish the energy scale of the single particle excitation across the CDW gap. Moreover, we find that the CDW collective state gaps a very large portion of the Fermi surface. Similarly to the related rare earth tri-tellurides, we envisage that interactions and Umklapp processes play a role in the onset of the CDW broken symmetry ground state.
\end{abstract}



\maketitle

\section{Introduction}
Charge-Density-Wave (CDW) is a paramount example of broken-symmetry ground state principally driven by the electron-phonon interaction and occurring in low dimensional materials. More than 50 years ago Peierls first pointed out that a one-dimensional (1D) metal coupled to the underlying lattice is indeed not stable at low temperatures \cite{peierls}. In most cases, the instability in 1D leads to a metal-insulator phase transition. The excitation spectrum of the resulting ground state of such a coupled electron-phonon system is characterized by a gap absorption feature, as a consequence of the Fermi surface (FS) nesting with wave vector $q=2k_{F}$, and by a collective mode formed by electron-hole pairs at $q=2k_{F}$ \cite{grunercdw}. In contrast to superconductors, the phase excitations of the collective mode are gapless. Consequently, electrostatic potentials (due to impurities, grain boundaries, surface effect etc.) break the translational symmetry and lead to the pinning of the collective mode. This ultimately results in a non conducting but highly polarizable ground state in 1D \cite{grunercdw}. By increasing the dimensionality of the interacting electron gas system the impact of the nesting is less relevant and the fraction of gapped FS is reduced. CDW transitions then lead in 2D to a pseudogap feature in the density of states and those 2D systems remain metallic down to low temperatures. That increasing the dimensionality less affects the FS is also well represented by the progressive suppression of the singularity in the Lindhard response function at $q=2k_{F}$, which in fact totally disappears in 3D \cite{grunercdw}. Furthermore, low dimensionality has a strong impact in the normal state properties, as well, which particularly in one dimension can be suitably explained within the Tomonaga-Luttinger liquid or Luther-Emery scenarios \cite{degiorgibook,vescoli}.

The formation of CDW ground states is by now well documented in a broad range of low dimensional solids \cite{grunercdw}, among which we encounter the linear chain organic and inorganic compounds. An effective low-dimensionality has been also identified in the layered $R$Te$_2$ ($R$=La and Ce), which play host to a CDW state, described in terms of modulated Cu$_2$ Sb-type structure (Pa/nmm) based on alternating layers of square-planar Te sheets and a corrugated $R$Te slab \cite{shin,kang}. It is worth noting that no clear electronic phase transitions are observed below 300 K, and the material appears to be deep in the CDW state even at room temperature. A substantial anisotropy in the electrical resistivity confirms the quasi-2D character of the charge carriers. The temperature dependence of the resistivity is moreover reminiscent of either a doped small-gap semiconductor or possibly a semimetal and is certainly far from that of a good metal \cite{shin}. A superlattice modulation of the average structure has been observed via transmission electron microscopy (TEM) \cite{shin,dimasi} and x-ray diffraction \cite{stowe}, with somewhat different modulation wave vectors for the two compounds. Tunneling experiment \cite{jung} and angle resolved photoemission spectroscopy (ARPES) reveal furthermore a large CDW gap (the ARPES gap extends between 100 and 600 meV \cite{shin,kang,XPSgap}). The gap was found to vary in magnitude around FS rather differently for the two compounds \cite{shin,kang}, presumably reflecting the different lattice modulations. These observations essentially establish the lattice modulation in $R$Te$_2$ as a CDW, driven by an electronic instability of FS. Since the title compounds are essentially quasi-2D materials, the FS is only partially gapped by the CDW transition, so that ungapped charge carriers contribute to the conductivity at low temperatures. Heat capacity measurements at low temperatures also indicate a very small electronic density of states, consistent with the measured electrical resistivity \cite{shin}.   

$R$Te$_2$ share several common features and similar properties with the related bilayer $R$Te$_3$ materials \cite{sacchettiprb}. The structural and electronic simplicity of $R$Te$_2$ and $R$Te$_3$, combined with the large size of the CDW gap, makes these materials particularly attractive for studying the effect of CDW formation on the electronic structure of layered systems. One approach is to exploit optical spectroscopy methods, which are ideal experimental tools to get insight into the absorption spectrum of the investigated materials. Optical data allow in principle to address both the gapped as well as the ungapped fractions of FS \cite{sacchettiprb}. We seek to identify and to discuss the gap excitation and the fraction of FS involved in the CDW transition. These measurements were also partially inspired by the desire to reconcile ARPES results (which show ungapped sections of FS \cite{shin,kang,XPSgap}) with dc resistivity
measurements (typical for a bad metal \cite{shin}). Furthermore, the frequency dependence of the absorption spectrum will shed light on the possible non-Fermi liquid nature of the electronic properties of these low dimensional layered compounds.

\section{Experiment and Results}

We report the first comprehensive optical study on LaTe$_2$ and CeTe$_2$ single crystals, which were grown in this case by slow cooling a
binary melt, as described previously in Ref. \onlinecite{shin}. Samples grown by this
technique are as close to stoichiometric as possible, having compositions
CeTe$_{2.00}$ and LaTe$_{1.95}$ as determined by electron microprobe analysis using
elemental standards, with an uncertainty of +/- 0.03 in the Te content. The crystals were polished in order to achieve a clean surface for the reflectivity measurements. Exploiting several spectrometers and interferometers, the optical reflectivity $R(\omega)$ was measured for all samples from the far-infrared (6 meV) up to the ultraviolet (6 eV) spectral range, with light polarized parallel to the Te planes. Details pertaining to the experiments can be found elsewhere \cite{sacchettiprb,dressel, wooten}.

\begin{figure}[!tb]
\center
\includegraphics[width=8.5cm]{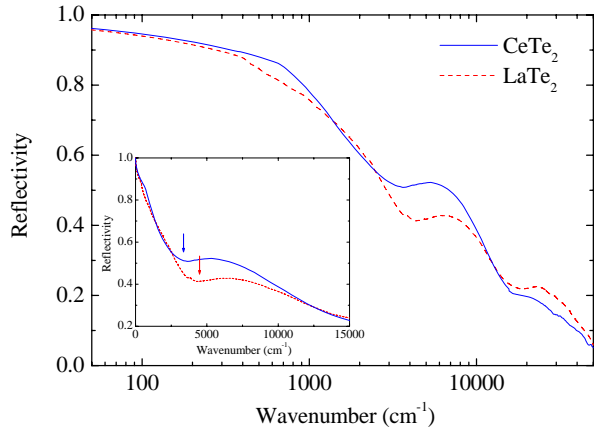}

\caption{$R(\omega)$ of LaTe$_2$ and CeTe$_2$ at 300 K. The inset is a blow-up of $R(\omega)$ emphasizing the broad mid infrared absorption feature in the spectral range around 6000 cm$^{-1}$. The arrows mark the shallow minimum in $R(\omega)$ at the onset of the broad absorption, peaked between 5000 and 6000 cm$^{-1}$ and overlapped to the plasma edge feature.}
\label{Refl}
\end{figure}

Figure 1 displays $R(\omega)$ for both title compounds over the whole investigated spectral range. $R(\omega)$ does not display any temperature dependence between 300 and 10 K in the covered energy interval and merges to total reflection for frequency going to zero for both compounds. The $R(\omega)$ spectra are therefore typical of an overall metallic behaviour. Overlapped to the plasma edge feature (for $\omega< 10^4$ cm$^{-1}$) one can observe furthermore a rather broad absorption peaked at about 6000 cm$^{-1}$. This feature is emphasized by the blow-up of $R(\omega)$ (inset Fig. 1) in the spectral range around 6000 cm$^{-1}$.

\begin{figure}[!tb]
\center
\includegraphics[width=8.5cm]{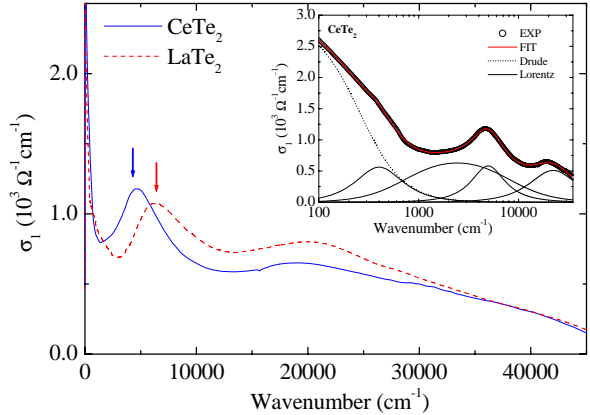}
\caption{Room temperature $\sigma_1(\omega)$ of LaTe$_2$ and CeTe$_2$. Arrows mark the position of the mid infrared peak for each compound (see text). Inset: Drude-Lorentz fit for
CeTe$_2$, showing the experimental data, the fitted curve, the
Drude and the Lorentz components.} \label{Sig1}
\end{figure}

These characteristic fingerprints of the electrodynamic response in $R$Te$_2$ are even better highlighted in the real part $\sigma_1(\omega)$ of the optical conductivity, shown in Fig. 2. $\sigma_1(\omega)$ is achieved through standard Kramers-Kronig transformation \cite{dressel, wooten} of $R(\omega)$. To this purpose $R(\omega)$  was extended towards zero frequency (i.e., $\omega\rightarrow 0$) with the Hagen-Rubens extrapolation $R(\omega$)=$1-2\sqrt{\omega/\sigma_{dc}}$ and with standard power-law extrapolations at high frequencies (i.e., $\omega>5$ x10$^{4}$ cm$^{-1}$). The dc conductivity values employed in the Hagen-Rubens extension of $R(\omega)$  are in fair agreement with transport measurements \cite{shin}. In $\sigma_1(\omega)$ we observe the so-called effective Drude resonance below 1000 cm$^{-1}$, due to the free-charge carriers, and the mid infrared absorption at 5000 and 6000 cm$^{-1}$ for CeTe$_2$ and LaTe$_2$ (arrows in Fig. 2), respectively, which will be later ascribed to the single particle excitation across the CDW gap. At higher frequencies there is also a broad feature centered at about 2x10$^{4}$ cm$^{-1}$ signaling the onset of electronic interband transitions. Tight binding band structure calculations reveal indeed the presence of excitations due to the electronic interband transitions well above 1 eV \cite{shin}.

In order to facilitate the discussion and for better identifying the characteristic features in the electrodynamic response of these rare-earth di-tellurides, we describe their absorption spectrum within the Lorentz-Drude phenomenological approach \cite{dressel,wooten}. It consists in reproducing the dielectric
function by the following expression: 
{\setlength\arraycolsep{2pt}
\begin{eqnarray}
\nonumber \tilde{\epsilon}(\omega) & = & \epsilon_1(\omega)
+i\epsilon_2(\omega) =
\\ & = & \epsilon_{\infty}-\frac{\omega_p^2}{\omega^2+i \omega
\gamma_D}+\sum_j \frac{S_j^2}{\omega^2-\omega_j^2-i \omega
\gamma_j},
\end{eqnarray}}
where $\epsilon_{\infty}$ is the optical dielectric constant,
$\omega_p$ and $\gamma_D$ are the plasma frequency and the width
of the Drude peak, whereas $\omega_j$, $\gamma_j$, and $S^2_j$ are
the center-peak frequency, the width, and the mode strength for
the $j$-th Lorentz harmonic oscillator (h.o.), respectively.
$\sigma_1(\omega)$ is then obtained from $\sigma_1(\omega)=\omega
\epsilon_2(\omega)/4\pi$.

The fits of all spectra were extended up to 4x10$^{4}$ cm$^{-1}$. Besides the Drude contribution, four Lorentz h.o.'s for each compound are required to fit all features at finite frequencies. The three lowest frequency h.o.'s will be associated below with the CDW gap excitation, while the fourth one will represent the first excitation at the onset of the high frequency electronic transitions. This is explicitly shown in the inset of Fig. 2 for the Ce-compound, where the fit components are displayed together with the total fit. The fit quality is astonishingly good for both investigated compounds.

\section{Discussion}
The depletion in the $\sigma_1(\omega)$ spectrum (Fig. 2) between the Drude component and the absorption feature at about 6000 cm$^{-1}$ bears a rather striking similarity with the situation encountered in the absorption spectrum of the bilayer $R$Te$_3$ \cite{sacchettiprb}. Therefore making use of the analogous notation of Ref. \onlinecite{sacchettiprb}, we claim that such a depletion in $\sigma_1(\omega)$ identifies the onset for the excitation across the CDW gap into the single particle ($SP$) states, giving rise to the mid infrared absorption. From now on we will refer to this mid infrared absorption as $SP$ peak. We note from Fig. 2 and its inset that the $SP$ peak coincides with the third Lorentz h.o. at about 5000 and 6000 cm$^{-1}$ for CeTe$_2$ and LaTe$_2$, respectively. Nonetheless, this $SP$ absorption feature is also characterized by a rather broad low frequency tail (represented in the fit by the two lowest h.o.'s), which extends well within the high frequency tail of the Drude component. This might indicate, on the one hand, that the metallic component is not simply Drude-like and that some localization occurs, or on the other hand, that the $SP$ peak in each compound can be thought as composed by the superposition of several excitations. These latter excitations, in analogy to $R$Te$_3$ \cite{sacchettiprb}, would mimic a continuous distribution of gap values, as seen in ARPES \cite{shin,kang}. This would also mean that the CDW gap differently affects the Fermi surface (FS), with perfectly nested regions of FS with a large gap and non-perfectly nested ones with a small gap.

Following our previous work on $R$Te$_3$ \cite{sacchettiprb} and tending towards the scenario in which the data are described by a distribution of gaps, we can similarly introduce a so-called averaged quantity $\omega_{SP}$:
\begin{equation}
\omega_{SP}=\frac{\sum_{j=1}^3 \omega_j S_j^2}{\sum_{j=1}^3
S_j^2},
\end{equation}
which represents the center of mass of the $SP$ excitation. We believe that $\omega_{SP}$ provides an average single energy scale, estimating the optical CDW gap. Table I displays the values of $\omega_{SP}$ for both compounds. $\omega_{SP}$ for the rare-earth di-tellurides decreases by almost a factor of two going from La to Ce, while it was almost identical between CeTe$_3$ and LaTe$_3$. $\omega_{SP}$ in LaTe$_2$ is comparable to the value for LaTe$_3$ and in CeTe$_2$ is smaller than in CeTe$_3$ \cite{sacchettiprb}. Our gap values in $R$Te$_2$ lie in the interval of CDW gaps, estimated from ARPES data \cite{shin,kang}.

The spectral weight encountered in the Drude component is also of interest: the plasma frequency $\omega_p$ in the CDW state remains almost constant (Table I) in the La and Ce di-tellurides, while it slightly increases from LaTe$_3$ to CeTe$_3$. $\omega_p$ obviously refers to the ungapped fraction of FS, thus responsible for the effective metallic behaviour even within the CDW phase. For these ungapped carriers in the CDW state, we assume a 2D free-holes scenario. The Sommerfeld $\gamma$-value of the specific heat and the charge carriers effective mass $m_{CDW}$ in the CDW phase are related by \cite{sacchettiprb}:
\begin{equation}
\gamma=\frac{\pi k_B^2 m_{CDW} a^2}{3 \hbar^2}. \label{mCDW}
\end{equation}
Using $\gamma=0.0003$ J/molK$^2$ for LaTe$_2$ \cite{shin,notegamma}, we achieve $m_{CDW}$ for both compounds, as reported in Table I. We can now exploit $\omega_p$ in order to obtain the free charge carriers concentration in the CDW state, listed in Table I, as well. 

\begin{table}[!t]
\centering
\begin{tabular}{|l|c|c|c|c|c|c|c|}
\hline
\textbf{Samples}&$\omega_{p}$&$\omega_{SP}$&$\Phi$&$m_{NS}$&$n_{NS}$&$m_{CDW}$&$n_{CDW}$\\
\hline
\textbf{LaTe$_2$} & 6840 & 5502 & 0.0845 & 1.2&1.3&0.3&0.03 \\
\hline
\textbf{CeTe$_2$}  & 6658 & 3165& 0.0805 & 1.2 & 1.4 & 0.3 & 0.03 \\
\hline
\end{tabular}
\caption{The first two columns summarize the values $\omega_p$ and $\omega_{SP}$ (eq. (2)), both in cm$^{-1}$. The third column reports the ratio
$\Phi$ between the Drude spectral weights (see text) in the CDW
and normal state. The last four columns report the effective mass $m_{NS}$ and carriers' density $n_{NS}$ in
the normal state and effective mass $m_{CDW}$ and carriers'
density $n_{CDW}$ in the CDW state for both samples. Carriers'
densities and effective masses are given in carriers/cell and
$m_e$ units, respectively.} \label{Tab}
\end{table}

It is now worth extending this analysis to the hypothetical normal state (NS), which should develop at temperatures much higher than 300 K. Again we make use of our previous analysis for the rare earth tri-tellurides \cite{sacchettiprb} and we exploit an elementary 2D tight-binding calculation involving the Te p$_x$ and p$_y$ orbitals \cite{shin}. The Fermi energy, associated to the holelike charge-carriers, is $E_F=1.95$ eV. Therefore, considering a parabolic expansion of the 2D bands around their maximum, $E_F$ is then related to the effective mass $m_{NS}$ in the (hypothetical) NS by \cite{sacchettiprb}:
\begin{equation}
E_F=\frac{\pi \hbar^2 n_{2D}}{m_{NS}}, \label{mNS}
\end{equation}
where $n_{2D}$ is the 2D hole density in NS. With $n_{2D}=2/a^2$ ($a$=4.5 \AA, for both compounds \cite{LattConst}), the resulting $m_{NS}$ are reported in Table I. $m_{CDW}$ and $m_{NS}$ describe the average curvature of the free charge carriers bands in CDW and NS, respectively. There is approximately a factor of four between $m_{NS}$ and $m_{CDW}$ (in  the $R$Te$_3$ compounds it was a factor of 2 \cite{sacchettiprb}). Such a difference suggests again that in the CDW phase the average curvature of the bands is larger than in NS.

The $m_{NS}$ values allow us to extract the NS charge carrier's concentration $n_{NS}$, which can be then directly compared with $n_{ch}$, estimated from the chemical counting \cite{kang}. Prerequisite for the achievement of $n_{NS}$ is first of all the estimation of the total spectral weight associated to the itinerant charge carriers in NS. If we assume the conservation of the
spectral weight between the CDW and the normal state, and that
there will be no $SP$ peak in the (hypothetical) NS, the Drude
contribution in NS would then have a total spectral weight given
by $S^2_{NS}=\omega^2_p+\sum_{j=1}^3 S_j^2$ \cite{sacchettiprb}. Therefore, from
$S_{NS}^2\sim n_{NS}/m_{NS}$ we obtain $n_{NS}$ as reported in Table I. The $n_{NS}$ values are indeed very close to $n_{ch}$=2 \cite{nchem}. While differences in
the stoichiometry of the two compounds due to Te vacancies might account
for a small part of this difference, the origin of the discrepancy most
likely lies in the approximations we have used to extract $n_{NS}$. Given the experimental uncertainties, the rather fair agreement between $n_{NS}$
and $n_{ch}$ adds confidence to the reliability of our analysis.

From our data we can also estimate the quantity $\Phi=\omega_p^2/(\omega_p^2+\sum_{j=1}^3 S_j^2)$ (Table I), representing the ratio between the spectral weight of the Drude peak and the total spectral weight of the Drude term and mid infrared absorption. $\Phi$ roughly measures, as previously shown in Ref. \onlinecite{sacchettiprb}, the fraction of the ungapped FS area (i.e., those parts of FS which are not affected by the CDW state). Since $\Phi$ is proportional to ($m_{NS}/m_{CDW})(n_{CDW}/n_{NS}$), it turns out to be about a factor four larger than the carriers' ratio in both systems (Table I). The very small values of $\Phi$, however comparable to those of $R$Te$_3$ \cite{sacchettiprb}, confirm that a large (and almost equal) portion of FS for both La and Ce di-tellurides is gapped. Large gapped regions of FS has been also evinced from recent ARPES data, even though the relative ratio of the gapped regions was found to be rather different for the two compounds \cite{shin}. The presence of only a very small numbers of carriers at the Fermi energy would explain the bad metallicity observed by electrical conductivity as well as heat capacity \cite{shin}. We also note that there is a substantial disorder in the La compound arising from the Te vacancies, an effect all but absent in the stoichiometric tri-telluride compounds. Therefore, polaronic and/or localization effects may also play a significant role in substantially reducing the conductivity.

\begin{figure}[!t]
\centering
\includegraphics[width=8.5cm]{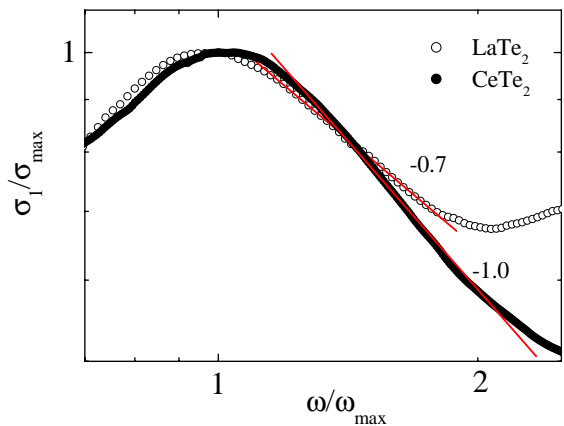}
\caption{$\sigma_1(\omega)$ of $R$Te$_2$ ($R$=La and Ce) plotted on a bi-logarithmic scale. The y-axis is scaled by the maximum of the mid infrared peak in $\sigma_1$, while the energy axis is scaled by the frequency where the maximum in $\sigma_1(\omega)$ occurs. The solid lines are power-law fits to the
data (the exponents are given in the figure).} \label{Power}
\end{figure}

We have also observed a power-law behaviour ($\sigma_1(\omega)\sim\omega^{\eta}$) in $R$Te$_2$ for frequencies above $\omega_{SP}$. This is shown in Fig. 3 for both investigated compounds, using a scaled representation. $\eta$ for $R$Te$_2$ ranges between -0.7 and -1 for the La and Ce compounds, respectively. These $\eta$ values compare rather well with those in $R$Te$_3$ and the power-law behaviour extends over an equivalent spectral range, as well \cite{sacchettiprb}. Such a power-law behaviour may be a further evidence for a typical Tomonaga-Luttinger liquid scenario, and emphasizes a non-negligible contribution of 1D correlation effects in the physics of these 2D compounds \cite{vescoli,sacchettiprb,giamarchi}. An exponent $\eta$ close to -1 could suggest a so-called antiadiabatic limit in the sense that one could integrate over the phonon field. The phonon fluctuations would then introduce an effective interaction between the particles. However, this scenario would be extremely unlikely since one would need characteristic phonon modes above 1 eV \cite{sacchettiprb,giamarchi}. Alternatively an exponent $\eta$ of the order of -1 could be reconciled with the theory if one assumes direct interaction between electrons, as source for Umklapp scattering \cite{giamarchi}. As shown already for $R$Te$_3$ \cite{sacchettiprb}, this theoretical framework would then be in reasonable agreement with the observed experimental behaviour in $R$Te$_2$, as well. Consequently a standard electron-phonon mechanism in $R$Te$_2$, as well as in $R$Te$_3$, may not be enough to fully account for the CDW formation.

Finally, we note that the low frequency side of $\sigma_1(\omega)$ below the $SP$ peak also possibly hints to a power-law behaviour with $\eta<1$ (not shown). However, such an exponent  $\eta$ in $R$Te$_2$ differentiates quite substantially from the values encountered in the Ce and La tri-tellurides \cite{sacchettiprb}. In these latter compounds $\sigma_1(\omega)\sim\omega^3$ at $\omega<\omega_{SP}$, which has been considered as a further clear indication for Umklapp processes; the dominant source of scattering underlying a 1D Mott (insulating) state \cite{giamarchi}. Any attempt to elucidate such a puzzling difference in the low frequency power-law behaviour between di- and tri-tellurides would be rather speculative at this point, since the observed low frequency power-law in $R$Te$_2$ develops over too small of a spectral range.

\section{Conclusions}
In summary, we have described the optical response in $R$Te$_2$ for $R$=La and Ce. Our data, besides revealing the $SP$ excitation across the CDW gap, also indicate that a large portion of FS is gapped for both compounds. This strengthens the view of the rare-earth di-tellurides as poor metals with small density of states at the Fermi level. From the high frequency power-law behaviour we evinced a hidden one-dimensional behaviour in these quasi-2D layered compounds, similar to that which is observed for the closely related $R$Te$_3$ compounds.

ARPES and thermodynamic experiments have pointed out the sensitive role played by perturbation, like Te vacancies, on the CDW state \cite{shin}. Such a sensitivity is even stronger than in $R$Te$_3$ and it has been mainly associated with the observation that the nesting wave vectors, particularly in CeTe$_2$, are somewhat poorly defined \cite{shin}. Besides the Te deficiencies, changes in the rare earth (chemical pressure) may also lead to subtle differences in the lattice modulation. Therefore, applied pressure might also affect to some extent the electronic structure and the CDW condensate and might therefore be used to differentiate between the
effects of chemical pressure and vacancy concentration in determining the
differences between these two compounds. Analogous to $R$Te$_3$ \cite{sacchettipressure} pressure dependent optical investigations may be of great relevance. Such investigations are in progress and will be reported elsewhere.

\begin{acknowledgments}
The authors wish to thank J. M\"uller for technical help and V. Brouet for fruitful discussions. One of us (A.S.) wishes to acknowledge financial support from the Della Riccia Foundation. This work has been supported
by the Swiss National Foundation for the Scientific Research
within the NCCR MaNEP pool. This work is also supported by the
Department of Energy, Office of Basic Energy Sciences under
contract DE-AC02-76SF00515.
\end{acknowledgments}


\begin{thebibliography}{99}
\bibitem{XPSgap} The gap values determined from the ARPES experiments are always
referred with respect to the Fermi energy. Assuming the Fermi energy in the
middle of the gap, means that the optically determined gap is about twice as
large as the ARPES one.
\bibitem{notegamma} Analogous  to $R$Te$_3$ \cite{sacchettiprb}, we use the $\gamma$ value of LaTe$_2$ for CeTe$_2$, as well. For magnetic members of the rare-earth series is difficult to extract $\gamma$ from thermodynamic experiments. However, since the chemical and structural changes across the series are small and even though the modulation wave vectors and density of states at $E_F$ are rather different among the two compounds \cite{shin,kang}, the assumption of equal $\gamma$ for both compounds seems at this point to be a reasonable approximation.
\bibitem{nchem} The ionic configuration of $R$Te$_2$ is considered to be $R^{3+}$Te(2)$^{2-}$Te(1)$^{1-}$ so as to produce hole carriers in the Te(1) sheets, precisely one hole for each formula unit.  Since the unit-cell contains 2 formula unit, the number of the hole carriers ($n_{ch}$) inside the unit cell is 2, as the number of  Te(1)$^{1-}$ atoms in the planar sheets \cite{kang}. We note that $n_{ch}$ corresponds to a 3/4 full band, since the p$_z$ orbitals are filled.
\end{thebibliography}
\end{document}